\documentclass[prl,aps,showpacs,twocolumn]{revtex4-1}

\usepackage{amsmath,amsfonts}

\usepackage{bm}

\usepackage{graphicx}
\usepackage[caption=false]{subfig}

\usepackage[squaren]{SIunits}
\usepackage{nicefrac}

\usepackage{hyperref}

\usepackage{color}

\newcommand{\define}{ \stackrel{\text{\tiny def.}}{=} }
\newcommand{\pdiff}[2]{ \ensuremath{ \frac{\partial #1}{\partial #2} } }
\newcommand{\form}[1]{ \,\mathrm{d} #1 \;}
\newcommand*{\im}{\ensuremath{i}}
\renewcommand{\vec}[1]{ \ensuremath{\bm{ #1 }}}

\begin{document}

\title{Focused Crossed Andreev Reflection}

\author{H\aa{}vard Haugen}%
\author{Arne Brataas}%
\affiliation{Department of Physics, %
  Norwegian University of Science and Technology, %
  N-7491 Trondheim, Norway}

\author{Xavier Waintal}%
\affiliation{SPSMS-INAC-CEA, %
  17 rue des Martyrs, %
  38054 Grenoble CEDEX 9, %
  France}

\author{Gerrit E. W. Bauer}%
\affiliation{Kavli Institute of NanoScience, %
  Delft University of Technology, %
  2628 CJ Delft, %
  The Netherlands}


\date{\today}

\begin{abstract}
  We consider non-local transport in a system with one superconducting
  and two normal metal terminals. Electron focusing by weak
  perpendicular magnetic fields is shown to tune the ratio between
  crossed Andreev reflection (CAR) and electron transfer (ET) in the
  non-local current response. Additionally, electron focusing
  facilitates non-local signals between normal metal contacts where the
  separation is as large as the mean free path rather
  than being limited by the coherence length of the
  superconductor. CAR and ET can be selectively enhanced by modulating
  the magnetic field.
\end{abstract}

\maketitle
Andreev reflection (AR) is a signature sub-gap scattering phenomena at
normal-superconductor (NS) interfaces. Two electrons (at energies
symmetrically around the chemical potential of the superconductor)
enter the superconducting condensate as a Cooper pair, resulting in a
retro-reflected hole on the normal side of the interface. The
superconducting coherence length $\xi$ determines the spatial extent
of the Cooper pairs and therefore gives the scale of the largest
possible separation between the incoming electron and the
retro-reflected hole (at the interface).

When two normal metal contacts $N_1$ and $N_2$ separated by a distance
$L \le \xi$ are connected to a superconductor, the Andreev reflected
holes arising from incoming electrons in $N_1$ also have a finite
probability of leaving the structure through $N_2$
\cite{byers:prl:v74:p306:y1995, deutscher:apl:v76:p487:y2000,
  falci:epl:v54:p255:y2001, feinberg:epjb:v36:p419:y2003}. This
non-local process, called crossed Andreev reflection (CAR), creates a
spatially separated phase-coherent electron-hole pair, and is a
candidate for a solid state entangler
\cite{recher:prb:v63:p165314:y2001}. Competing with CAR is a process
called electron transfer (ET), in which an electron propagates from
$N_1$ to $N_2$ either directly or via a virtual excitation in the
superconductor \cite{deutscher:apl:v76:p487:y2000}. The second
process, involving a virtual excitation, is also referred to as
electron co-tunneling.
The competition between CAR and ET has been studied for about a decade
\cite{%
  byers:prl:v74:p306:y1995, %
  hartog:prl:v77:p4954:y1996, %
  deutscher:apl:v76:p487:y2000, %
  falci:epl:v54:p255:y2001, %
  recher:prb:v63:p165314:y2001, %
  beckmann:prl:v93:p197003:y2004, %
  russo:prl:v95:p027002:y2005, %
  morten:prb:v74:p214510:y2006, %
  *morten:prb:v78:p224515:y2008, %
  cadden-zimansky:prl:v97:p237003:y2006, %
  cadden-zimansky:nphys:v5:p393:y2009%
}. Theoretical papers report that ET typically dominates CAR in
linear response \cite{falci:epl:v54:p255:y2001,
  morten:prb:v74:p214510:y2006, morten:prb:v78:p224515:y2008}. CAR
can dominate ET beyond linear response or in the presence of
interactions \cite{russo:prl:v95:p027002:y2005,
  cadden-zimansky:prl:v97:p237003:y2006,
  levyyeyati:nphys:v3:p455:y2007, golubev:prl:v103:p067006:y2009}.

\begin{figure}
  \centering
  \subfloat[Focused crossed Andreev reflection]{
    \label{fig:focused:crossed:andreev:car}
    \includegraphics[width=0.9\columnwidth]{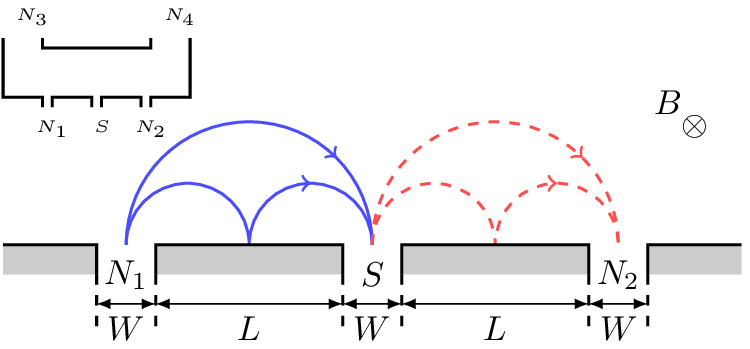}
  }%
  \\
  \subfloat[Focused electron transfer]{
    \label{fig:focused:crossed:andreev:et}
    \includegraphics[width=0.9\columnwidth]{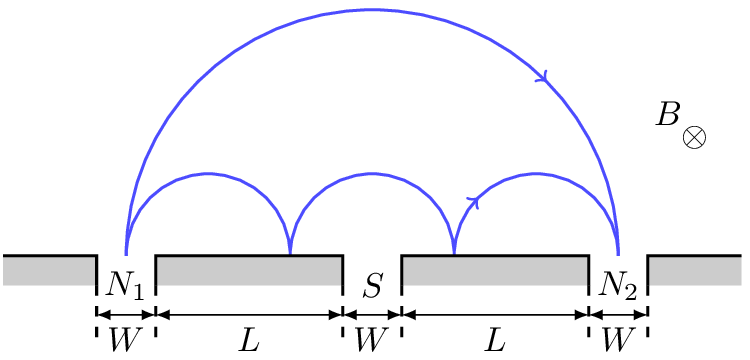}
  }%
  \caption{(Color online) Illustration of focused crossed Andreev
    reflection. %
    \protect\subref{fig:focused:crossed:andreev:car} %
    When the separation between $N_1$ and $N_2$ is an even integer
    multiple of the cyclotron diameter $d_c$, electron focusing
    enhances CAR and leads to a negative non-local conductance. %
    \protect\subref{fig:focused:crossed:andreev:et} %
    When the separation is an odd multiple of $d_c$, ET is enhanced
    and we expect a positive peak in the non-local conductance. Inset
    of \protect\subref{fig:focused:crossed:andreev:car}: Device used
    in the simulations. }
  \label{fig:focused:crossed:andreev}
\end{figure}

Much work on CAR has focused on the spin manipulation of the carriers,
with normal contact separation $L$ limited by the superconducting
coherence length $\xi$ \cite{deutscher:apl:v76:p487:y2000,
  recher:prb:v63:p165314:y2001, falci:epl:v54:p255:y2001,
  beckmann:prl:v93:p197003:y2004, cadden-zimansky:nphys:v5:p393:y2009,
  wei:nphys:v6:p494:y2010}.
In this Letter, we study electron and hole focusing by a weak
perpendicular magnetic field in a high-mobility two-dimensional
electron gas (2DEG) \cite{houten:prb:v39:p8556:y1989} We attach a
single superconducting contact between two normal contacts along the
edge of the device. In our scheme AR induces electron-hole
correlations on length scales which are only limited by the mean free
path $l_{\text{mf}}$ rather than $\xi$. For typical superconductors,
$\xi$ lies between $\unit{10}{\nano\metre}$ and
$\unit{100}{\nano\metre}$, \cite{levyyeyati:nphys:v3:p455:y2007} while
$l_{\text{mf}}$ can reach several microns in 2DEGs
\cite{beenakker:suplatt:v5:p127:y1989}
Very high mobilities have also been reported for graphene
\cite{bolotin:ssc:v146:p351:y2008, chen:nanolett:v9:p1621:y2009},
which is another candidate for focused CAR. The tuning between CAR and
ET through the external magnetic field is possible through the orbital
degrees of freedom at field strengths that do not introduce a spin
selectivity of the contacts. As long as the field is smaller than the
critical field of the superconductor, the relative magnitude of CAR
and ET can be controlled by varying the magnetic field.

The basic mechanism is illustrated in
Fig.~\ref{fig:focused:crossed:andreev}. An electron is injected from
the left contact $N_1$ by a small voltage bias.  For weak magnetic
fields, the motion of the electrons and holes can be understood in
terms of semi-classical cyclotron orbits
\cite{benistant:prl:v51:p817:y1983, houten:prb:v39:p8556:y1989} as a
result of the Lorentz force. For certain magnetic field strengths
(Fig.~\ref{fig:focused:crossed:andreev:car}), the electrons from $N_1$
are focused on the superconducting center contact $S$, at which an
Andreev reflected hole is emitted. Since AR changes the sign of both
charge and effective mass, the holes will feel the same Lorentz force
as the electrons and are therefore focused on contact $N_2$ to the
right of $S$ at the same distance as $N_1$ \cite{
  benistant:prl:v51:p817:y1983, tsoi:rmp:v71:p1641:y1999,
  giazotto:prb:v72:p054518:y2005}. The probability for CAR is thus
enhanced at the cost of ET. On the other hand, ET is enhanced at
certain other magnetic fields at which the incoming electrons are
focused on $N_1$, but where the skipping orbits do not interact with
the superconductor (Fig.~\ref{fig:focused:crossed:andreev:et}).

Andreev reflection in the presence of a magnetic field has been
thoroughly studied in the literature \cite{
  bozhko:pzetf:v36:p123:y1982, benistant:prl:v51:p817:y1983,
  tsoi:rmp:v71:p1641:y1999, giazotto:prb:v72:p054518:y2005,
  polinak:prb:v74:p132508:y2006, rakyta:prb:v76:p064516:y2007}.
Electron focusing has been used for the first direct observation of
Andreev reflection at an NS interface
\cite{bozhko:pzetf:v36:p123:y1982,
  benistant:prl:v51:p817:y1983}. Resonant enhancement of CAR due to
Andreev bound states at an NS interface has been proposed
\cite{polinak:prb:v74:p132508:y2006,
  rakyta:prb:v76:p064516:y2007}. Recently, an Andreev interferometer
was used to demonstrate the phase coherent nature of CAR and ET
\cite{cadden-zimansky:nphys:v5:p393:y2009}. We show here that electron
focusing clearly discriminates between CAR and ET, which might be
useful to maximize entanglement generation in artificial solid state
devices.

Before delving into the fully quantum mechanical treatment, we now
discuss the physics of electron focusing in a semi-classical
picture. The length scale associated with the semi-classical motion of
electrons with momentum $\hbar k_F$ in a magnetic field $B$ is the
cyclotron radius $r_c$ for electrons at the Fermi surface. As seen
from Fig.~\ref{fig:focused:crossed:andreev}, the natural parameter in
electron focusing is the cyclotron diameter
\begin{equation}
  \label{eq:cyclotronradius}
  \begin{split}
    d_c
    = \frac{2 \hbar k_F}{ e B}
    ,
  \end{split}
\end{equation}
where we assume high mobility, i.e. $d_c \ll l_{\text{mf}}$
\cite{houten:prb:v39:p8556:y1989}. Electron focusing between the
normal contacts in Fig.~\ref{fig:focused:crossed:andreev} occurs when
the distance $2 L$ between $N_1$ and $N_2$ is an integer multiple of
the cyclotron diameter,
$
  2 L 
  = n d_c
  ,
$
where $n$ is a positive integer. ET is enhanced for odd $n$, while CAR
will be enhanced for even $n$. The focusing field
\cite{houten:prb:v39:p8556:y1989},
\begin{equation}
  \label{eq:B:focus}
  B_\text{focus}
  =
  \frac{ 2 \hbar k_F }{e L}
  ,
\end{equation}
determines the magnetic field scale for which focusing features can be
expected. 

For strong magnetic fields the system enters the quantum Hall (QH)
regime, where the charge carriers are better described as chiral edge
states than semi-classical skipping orbits
\cite{beenakker:suplatt:v5:p127:y1989}. The characteristic length
scale associated with the QH regime is the magnetic length $l_B$,
which is the radius of the disc that encloses one flux quantum, $\pi
l_B^2 B = \Phi_0 = h/2e$. To avoid the QH regime, the magnetic flux
density $n_B = 1/(\pi l_B^2)$ must be substantially lower than the
electron density $n = k_F^2/(2\pi)$, or $B \ll \frac{h}{2 e} n \approx
\unit{7}{\tesla}$, for typical values for the electron density in a
2DEG, $n \approx \unit{3.5 \times 10^{15}}{\metre^{-2}}$
(corresponding to $\lambda_F \approx \unit{40}{\nano\meter}$)
\cite{houten:prb:v39:p8556:y1989}.  Superconductors with upper
critical fields above $\unit{10}{\tesla}$ are readily available
\cite{niu:prl:v91:p027002:y2003}. We expect CAR to be enhanced also in
the QH regime, since the edge states will be forced to interact with
the superconductor on the way from $N_1$ to $N_2$.

We will now confirm the semi-classical predictions 
by a numerical quantum calculation of the non-local
transport properties of the device shown in
Fig.~\ref{fig:focused:crossed:andreev}. The competition between CAR
and ET is studied through the non-local conductance
\cite{falci:epl:v54:p255:y2001, morten:prb:v74:p214510:y2006,
  morten:prb:v78:p224515:y2008},
\begin{equation}
  \label{eq:non-local:conductance}
  G_{21}
  \define
  - \frac{ \partial I_2 }{ \partial V_1 }
  =
  G_{21}^{\text{ET}}
  -
  G_{21}^{\text{CAR}}
  ,
\end{equation}
where $I_2$ is the current response in contact $N_2$ due to the
application of a voltage $V_1$ in the normal metal contact $N_1$ while
N2 and S are grounded. The overall minus sign is due to the definition
of the currents to be positive when electrons leave the
reservoirs. The difference in sign of $G_{21}^{\text{ET}}$ and
$G_{21}^{\text{CAR}}$ in Eq.~\eqref{eq:non-local:conductance} is due
to the fact that the outgoing current in $N_2$ produced by ET consists
of negatively charged electrons, while CAR contributes with positively
charged holes.

In our calculation we employ the standard 2DEG Hamiltonian
\begin{equation}
  \label{eq:2DEG:hamiltonian}
  \mathcal{H}(\vec{r}) = \frac{\vec{p}^2}{2 m} + V(\vec{r}) - \mu,
\end{equation}
where $\vec{p} = -\im \hbar \nabla + e \vec{A}(\vec{r})$ is the
canonical momentum and $m$ the effective
mass.
The Hamiltonian \eqref{eq:2DEG:hamiltonian} is extended it to Nambu
space \cite{nambu:pr:v117:p648:y1960}
\begin{equation}
  \label{eq:2DEG:nambu:hamiltonian}
  H 
  = 
  \int \form{\vec{r}}
  \Psi^\dagger(\vec{r}) 
  \begin{pmatrix}
    \mathcal{H}(\vec{r}) & \Delta(\vec{r}) \\
    \Delta^*(\vec{r}) & -\mathcal{H}^*(\vec{r})
  \end{pmatrix}
  \Psi(\vec{r})
  ,
\end{equation}
where at the contact $S$ the superconducting pair potential
$\Delta(\vec{r})$ is assumed to vary abruptly on the scale of the
Fermi wavelength $\lambda_F$, and is therefore modelled as step
function which is non-zero only inside the center contact $S$.  All
energies are measured from the chemical potential $\mu$ of the
superconductor.  The Nambu spinor $\Psi$ is defined in terms of the
field operators $\psi$ as
$\Psi = (\psi, \psi^\dagger)^T$.
A perpendicular magnetic field $\vec{B} = \nabla \times \vec{A} = B
\vec{e}_z$ is included everywhere except in the superconductor, which
expels the field \cite{hoppe:prl:v84:p1804:y2000}, and we consider
only elastic scattering.

At zero temperature, quantum interference due to scattering at the
sharp boundaries close to the contacts can mask the electron focusing
effect \cite{houten:prb:v39:p8556:y1989}. We therefore calculate the
non-local differential conductance at finite temperature, using the
standard formula,
\begin{equation}
  \label{eq:non-local:conductance:ft}
  G_{ij}
  =
  \int \form{\varepsilon} G_{ij}(\varepsilon)
  \left( - \pdiff{n_F(\varepsilon)}{\varepsilon} \right)
  ,
\end{equation}
where $n_F$ is the Fermi-Dirac distribution function.

We use the knitting algorithm presented in
Ref.~\onlinecite{kazymyrenko:prb:v77:p115119:y2008} to calculate the
self energies and retarded and advanced Green functions. Standard
expressions relate the conductance and current density to these
quantities. The device used in the simulations is sketched in the
inset of Fig.~\ref{fig:focused:crossed:andreev:car}, where the two
auxiliary contacts $N_3$ and $N_4$ are added to prevent
back-reflection of electrons into $N_1$. All edges 
cause specular electron scattering only.

\begin{figure}
  \centering
  \includegraphics[width=\columnwidth]{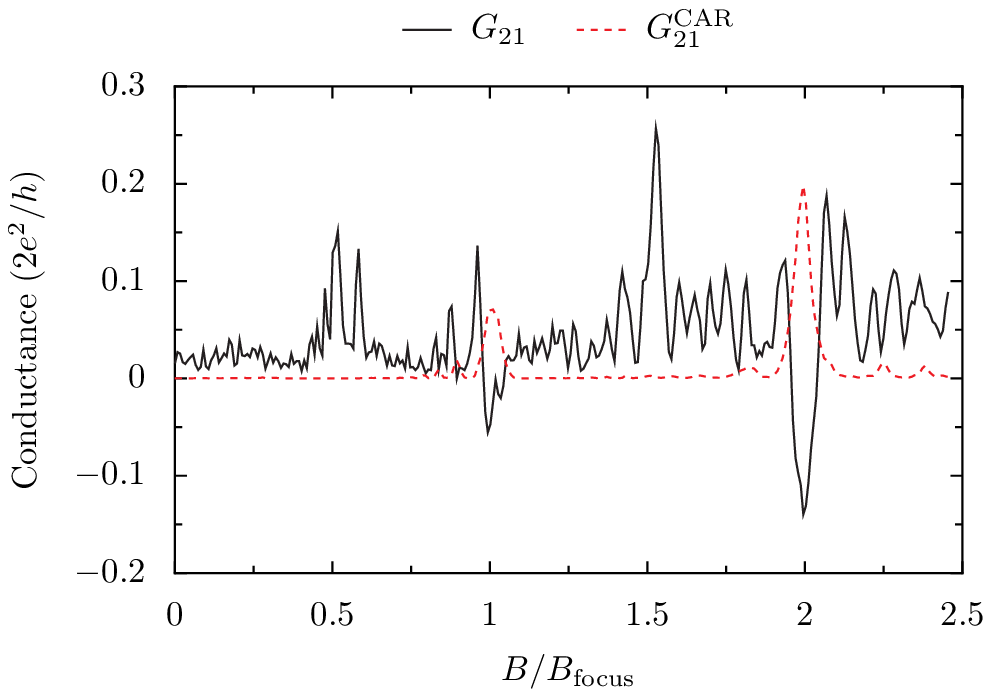}
  \caption{(Color online) Non-local conductance $G_{21}$ (solid black)
    for a device with point contacts ($W 
    \approx \lambda_F/2$). The magnetic field is given in units of the
    focusing field $B_{\text{focus}} = \unit{0.406}{\tesla}$. The
    contribution from $G_{21}^{\text{CAR}}$ (dashed red) shows that
    the negative peaks in $G_{21}$ are evidence for CAR at integer
    multiples of $B/B_{\text{focus}}$. }
  \label{fig:2deg:conductance:narrow:contacts}
\end{figure}

Figure~\ref{fig:2deg:conductance:narrow:contacts} shows the calculated
non-local conductance from Eq.~\eqref{eq:non-local:conductance:ft} as
a function of perpendicular magnetic field at a temperature of $T =
\unit{1}{\kelvin}$. The value chosen for the pair potential $\Delta$
would correspond to Pb, which has a critical temperature of $T_c
\approx \unit{7}{\kelvin} \gg T$ \cite{lide:y2010}. Also, since $T <
T_c/2$, we disregard the temperature dependence of the pair potential,
$\Delta(T) \approx \Delta(0)$ \cite{bardeen:pr:v108:p1175:y1957}.

The injector $N_1$, superconducting $S$, and collector $N_2$ contacts
are point contacts with width $W \approx \lambda_F/2$, so that only a
single mode contributes to the current
\cite{beenakker:suplatt:v5:p127:y1989}. The distance $L =
\unit{500}{\nano\metre}$ between the contacts corresponds to a
focusing field of $B_{\text{focus}} = \unit{(0.39 \pm 0.02)}{\tesla}$,
where the uncertainty is due to the finite width $W$ of the contacts
relative to $L$. The value found in the simulation agrees with the
expectation within the uncertainty dictated by the finite size of the
contacts.

The enhancement of CAR due to electron focusing is clearly seen in
terms of the two negative peaks in
Fig.~\ref{fig:2deg:conductance:narrow:contacts}. The position of these
peaks at integer values of $B/B_{\text{focus}}$ and the explicit
calculation of the contribution from $G_{21}^{\text{CAR}}$ in
Eq.~\eqref{eq:non-local:conductance} (dashed red line) is consistent
with the semi-classical interpretation presented earlier. The expected
enhancement of ET at half-integer $B/B_{\text{focus}}$ is somewhat
masked by interference peaks, but the positive peaks in $G_{21}$ when
$B/B_{\text{focus}}$ equals $1/2$ and $3/2$ are clearly visible. As
the magnetic field increases beyond $2.5 \, B_{\text{focus}}$
, the system gradually enters the QH regime, where
transport is associated with chiral edge states. 

\begin{figure}
  \centering
  \subfloat[Non-superconducting center contact ($\Delta = 0$)]{
    \label{fig:2deg:currdens:finite:temp:nodelta}
    \includegraphics[width=\columnwidth]{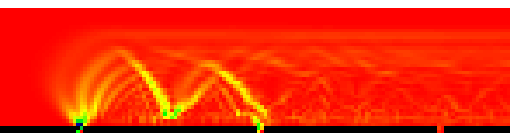}
  }
  \\
  \subfloat[Superconducting center contact ($\Delta \neq 0$)]{
    \label{fig:2deg:currdens:finite:temp:delta}
    \includegraphics[width=\columnwidth]{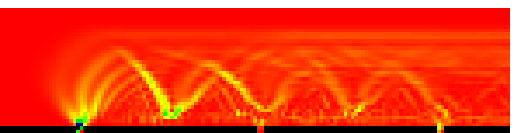}
  }
  \caption{(Color online) Electronic current density in a
    perpendicular magnetic field at $T = \unit{1}{\kelvin}$. Two
    skipping orbits, corresponding to a magnetic field $B \approx 2 f
    B_{\text{focus}}$, are clearly visible. %
    \protect\subref{fig:2deg:currdens:finite:temp:nodelta} %
    With a non-superconducting center contact $S$, a large portion of
    the current injected through contact $N_1$ leaves the structure
    through $S$. %
    \protect\subref{fig:2deg:currdens:finite:temp:delta} %
    When $S$ is superconducting, the Andreev reflected holes from $S$
    contribute to the current from $S$ to $N_2$.}
  \label{fig:2deg:currdens:finite:temp}
\end{figure}

The focusing enhancement of CAR at $B/B_{\text{focus}} = 1$ and $2$ in
Fig.~\ref{fig:2deg:conductance:narrow:contacts} can be visualized by
calculating the charge current density due to electrons injected from
contact $N_1$. This is shown in
Fig.~\ref{fig:2deg:currdens:finite:temp}, where we have set $B \approx
2 B_{\text{focus}}$. A skipping orbit between $N_1$ and $S$ is clearly
visible. Also visible is the diffraction of the incoming current
through $N_1$, which leads to a broadening of the skipping orbit
trajectories. In Fig.~\ref{fig:2deg:currdens:finite:temp:nodelta} the
center contact $S$ is normal ($\Delta = 0$). A large portion of the
injected current is then extracted through $S$.  In contrast, when $S$
is in the superconducting state, as shown in
Fig.~\ref{fig:2deg:currdens:finite:temp:delta}, the current density
increases substantially between $S$ and $N_2$ due to CAR.

\begin{figure}
  \centering
  \includegraphics[width=\columnwidth]{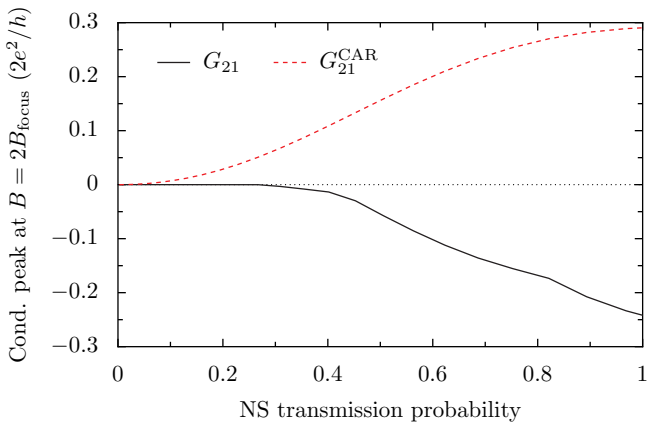}
  \caption{(Color online) Magnitude of peak in $G_{21}$ and 
    $G_{21}^{\text{CAR}}$ at $B/B_{\text{focus}} = 2$ as a function
    of transmission probability of
    the interface between the 2DEG and the superconductor. The
    conductance is calculated at zero temperature. 
  }
  \label{fig:NS:interface:transmission:probability}
\end{figure}

The technology to manufacture good contacts between superconducting
metals and 2DEGs has been developed for several types of
heterostructures \cite{takayanagi:prl:v54:p2449:y1985,
  boulay:jap:v105:p123919:y2009}. Although experimentally challenging
due to the presence of important Schottky barriers, fairly high
transparencies have been reported (for instance transmission
probability $\sim0.55$ with a critical field of 2T in the In-GaAs
heterostructures interfaces presented in
Ref.~\onlinecite{boulay:jap:v105:p123919:y2009}). In
Fig.~\ref{fig:NS:interface:transmission:probability}, we show the
height of the CAR peak at $B/B_{\text{focus}} = 2$ (for $T = 0$) as a
function of the transmission probability of the NS contact. The CAR
peak diminishes with decreasing quality of the interface but not
dramatically so.  We conclude that the effect should 
be observable with the available technology.

In conclusion, we have shown that electron focusing can be used to
enhance CAR over length scales much larger than the superconducting
coherence length $\xi$. The limiting length scale for electron
focusing and therefore CAR enhancement, is the mean free path
$l_{\text{mf}}$, which can be several orders of magnitude larger than
$\xi$ \cite{houten:prb:v39:p8556:y1989,
  levyyeyati:nphys:v3:p455:y2007}.  CAR is enhanced at the cost of ET
for magnetic fields that are integer multiples of the focusing field
in Eq.~\eqref{eq:B:focus}. At half integer multiples of the focusing
field, CAR plays a negligible role since the electron orbits avoid
the superconducting contact. Instead ET is enhanced as in normal
electron focusing \cite{houten:prb:v39:p8556:y1989}. %
The necessary magnetic field is relatively weak, and should be an
easily accessible experimental ``knob'' for controlling the CAR
enhancement.

CAR has been proposed as a means to create a solid state entangler,
using the natural entanglement of Cooper pairs. However, in most
systems quasiparticle backscattering into the injector contacts is a
serious limitation \cite{morten:epl:v81:p40002:y2008}. This difficulty
does not exist in our scheme.

\begin{acknowledgments}
  This work was supported by The Research Council of Norway through
  grant no. 167498/V30.
\end{acknowledgments}

\bibliography{caretfocus}

\end{document}